\documentclass[pre,twocolumn,twoside,superscriptaddress,nofootinbib]{revtex4}
\usepackage{graphicx,amssymb,amsmath,epsf}
\epsfclipon

\usepackage{xspace}
\usepackage{color}

\definecolor{antiquebrass}{rgb}{0.8, 0.58, 0.46}
\definecolor{mygreen}{rgb}{0.19, 0.73, 0.53}
\definecolor{bistre}{rgb}{0.24, 0.17, 0.12}

\newcommand{\be}{\begin{equation}}
\newcommand{\ee}{\end{equation}}

\newcommand{\co}{\text{c}}

\begin{document}

\title{Scale invariance implies conformal invariance for the three-dimensional Ising model}

\author{Bertrand Delamotte}
\author{Matthieu Tissier}
\affiliation{LPTMC, UPMC, CNRS UMR 7600, Sorbonne Universit\'es, 4 Place Jussieu, 75252 Paris Cedex 05, France}
\author{Nicol\'as Wschebor$^{1,}$}
\affiliation{Instituto de F\'{\i}sica, Faculdad de Ingenier\'{\i}a, 
Universidad de la Rep\'ublica, J.H. y Reissig 565, 11000 Montevideo, Uruguay}

\begin{abstract}
  Using Wilson renormalization group, we show that if no integrated
  vector operator of scaling dimension $-1$ exists, then scale
  invariance implies conformal invariance. By using the Lebowitz
  inequalities, we prove that this necessary condition is fulfilled in
  all dimensions for the Ising universality class. This shows, in
  particular, that scale invariance implies conformal invariance for
  the three-dimensional Ising model.
\end{abstract}
\date{\today}

\pacs{ }
\maketitle

\section{Introduction}

Conformal symmetry plays a considerable role both in high energy and
condensed matter physics. There has been  a renewed interest 
in recent years, particularly because of the AdS/CFT conjecture
\cite{Maldacena:1997re} and the successful use of conformal methods in
three-dimensional critical physics
\cite{ElShowk:2012ht,Gliozzi:2014jsa,El-Showk:2014dwa,Kos:2013tga,El-Showk:2013nia,Nakayama:2014sba}. The groundbreaking papers of
the 1970s and 1980s
\cite{Virasoro:1969zu,Polyakov:1970xd,Migdal:1972tk,Belavin:1984vu,Zamolodchikov:1986gt,Polchinski:1987dy}
solved two fundamental issues in two dimensions: First, scale
invariance implies conformal invariance under mild assumptions
\cite{Zamolodchikov:1986gt,Polchinski:1987dy} and, second, conformal
symmetry enables us to solve  most of the scale invariant problems 
that is, to determine critical exponents and correlation functions \cite{DiFrancesco:1997nk}.

An important ingredient
for the exact solution of two-dimensional conformal models
is the existence of an infinite number of generators
of the conformal group.
In higher dimensions, the number of
generators  is finite, and we could naively
conclude that symmetry arguments alone are not sufficient to solve a model
in the critical regime. However, it is well-known that
scale-invariant
theories are in a one-to-one correspondence with the fixed points of
the Wilson Renormalization Group (RG) \cite{Wilson:1973jj}, and that
the fixed point of a theory completely determines all the correlation
functions of a critical model for sufficiently small wavenumbers.
Therefore, at the level of principles, scale (and {\it a fortiori}
conformal) invariance is sufficient to determine all the universal critical
properties of a model.
Of course, in practice, the computation of these critical properties
requires us to solve the functional Wilson RG equations. This
is a formidable task that we do not know how to carry out without approximations.  Any supplementary information,
even if redundant, is therefore welcome and this is what conformal
invariance could provide. A breakthrough in this direction has been
achieved these last years with the conformal bootstrap program \cite{ElShowk:2012ht,Gliozzi:2014jsa,El-Showk:2014dwa}, which
 led to the {\it exact} (although numerical) computation of the
critical exponents of the Ising model in three dimensions assuming,
among other things, conformal invariance.

In parallel, a large activity has been devoted to understanding the
relation between scale and conformal invariance in -- or close to -- four
dimensions. It has been proven to all orders of
perturbation theory that scale invariance implies conformal invariance
\cite{Jack:1990eb} in
four-dimensional
unitary and Poincar\'e invariant
theories. Moreover, there are  strong indications that a
non-perturbative proof could be at reach in this dimension
\cite{Luty:2012ww,Dymarsky:2013pqa,Dymarsky:2014zja}.  

Despite decades of effort, it is still an open question to know
whether a typical statistical model is conformally invariant at
criticality in three dimensions. The aim of this article is
twofold. First, we derive a sufficient condition which, when
fulfilled, ensures that scale invariance implies conformal invariance.
In the second part of the paper, we prove that this condition is
fulfilled in any dimension for the euclidean $\mathbb{Z}_2$ model.

The rest of the paper is organized as follows. In
Sect.~\ref{sec_nprg}, we make a brief review of the nonperturbative
renormalization-group. We then recall in Sect.~\ref{sec_scale}
the relation between scale invariance and the NPRG. By using the same
methods, we generalize these considerations to the case of conformal
invariance in Sect. \ref{sec_conformal}. In
Sect. \ref{sec_sufficient}, we finally derive a sufficient condition
for the validity of conformal invariance in scale invariant models. We
show, on general arguments that this condition is expected to be
fulfilled in O($N$)  models (and in generalizations thereof). In
Sect~\ref{sec_proof}, these considerations are made rigourous for the
Ising universality class. We give our conclusions in
Sect~\ref{sec_conclusion}.

\section{Nonperturbative renormalization-group formalism}
\label{sec_nprg}

The proof of conformal invariance in all dimensions presented below is
intimately related to the deep structure of the Wilson RG
\footnote{The history of the relation between conformal invariance and
  Wilson RG is long, see for example
  \cite{Schafer:1976ss,Rosten:2014oja}.}  and scale invariance. We
therefore start by recalling, in the case of the $\mathbb{Z}_2$ model,
the formalism of the modern formulations (sometimes called
Nonperturbative RG, or functional RG) of the Wilson RG
\cite{Polchinski:1983gv,Wetterich:1992yh,Ellwanger:1993kk,Morris:1993qb,Berges}.
The coarse-graining procedure at some RG scale $k$ is implemented by
smoothly decoupling the long-wavelength modes
$\varphi(\vert q\vert<k)$ of the system, also called the slow modes,
by giving them a large mass, while keeping unchanged the dynamics of
the short-wavelength/rapid ones $\varphi(\vert q\vert>k)$.  This
decoupling is conveniently implemented by modifying the action or the
Hamiltonian of the model:
$S[\varphi]\to S[\varphi]+ \Delta S_k[\varphi]$ where
$\Delta S_k[\varphi]$ is quadratic in the field and reads, in Fourier
space,
$\Delta S_k[\varphi]=1/2 \int_q\, R_k(q^2)\varphi(q) \varphi(-q)$.
The precise shape of $R_k(q^2)$ does not matter for what follows as
long as it can be written as
\begin{equation}
\label{cutoff-RG}
 R_k(q^2) =Z_k k^2 r(q^2/k^2)
\end{equation}
where  $Z_k$ is the field renormalization factor and $r$ is a function that, (i)
falls off rapidly to 0 for $q^2>k^2$ -- the rapid modes $\varphi(\vert q\vert>k)$ are not affected by $\Delta S_k$ --
and (ii) goes to a constant for $q^2=0$ -- the slow modes $\varphi(\vert q\vert<k)$ acquire a mass of order $k$
and thus smoothly decouple. The
partition function, which now depends on the RG parameter $k$, reads:
\begin{equation}
\label{eq_part}
 {\cal Z}_k[J]=\int {\mathcal D}\varphi\, \exp\left(- S[\varphi]- \Delta S_k[\varphi] + \int_x J\varphi\right)
\end{equation}
where the field $J$ is a source term which corresponds to the magnetic field in the Ising model and
where the ultra-violet (UV) regime of the functional integral is assumed to be regularized at a momentum scale 
$\Lambda$, see for instance \cite{Machado:2010wi} for a lattice  
regularization in this formalism.
It is convenient to define the free energy
${\cal W}_k[J]= \ln{\cal Z}_k[J]$ and its (slightly modified) Legendre transform by
\begin{equation}
\label{eq_legendre}
 \Gamma_k[\phi]+{\cal W}_k[J]= \int_x J\phi -\frac 1 2 \int_{xy} R_k(\vert x-y\vert)\,\phi(x)\phi(y),
\end{equation}
with $\phi(x)=\delta {\cal W}_k/\delta J(x)$, $R_k(\vert x-y\vert)$ the
inverse Fourier transform of $ R_k(q^2)$ and where the last term has
been added for the following reason.  When $k$ is close to $\Lambda$, all
modes are completely frozen by the $\Delta S_k$ term because, for all
$q$, $R_{\Lambda}(q^2)$ is very large. Thus, ${\cal Z}_{k\to\Lambda}$
can be computed by the saddle-point method and it is then
straightforward to show that the presence of the last term in
Eq.~(\ref{eq_legendre}) leads to $\Gamma_{\Lambda}[\phi]\simeq
S[\varphi=\phi] $. On the contrary, when $k=0$, the definition of
$R_k$ implies that $R_{k=0}(q^2)\equiv0$ and the original model is
recovered: ${\cal Z}_{k=0}[J]= {\cal Z}[J]$ and
$\Gamma_{k=0}[\phi]=\Gamma[\phi]$, with $\Gamma[\phi]$ the usual Gibbs
free energy or generating functional of one-particle-irreducible
correlation functions. 

The exact RG equation for $\Gamma_k$ reads
\cite{Wetterich:1992yh,Ellwanger:1993kk,Morris:1993qb}:
\begin{equation}
\label{eqNPRG}
\partial_t \Gamma_k[\phi]=\frac 1 2 \int_{xy} \partial_t R_k(\vert x-y\vert)G_{k,x y}[\phi],
\end{equation}
where $t=\ln(k/\Lambda)$, and $G_{k,x y}[\phi]$ is the field-dependent propagator:
\begin{equation}
G_k=(\Gamma_{k}^{(2)}+R_k)^{-1}\ ,\ 
\Gamma_{k,xy}^{(2)}[\phi]= \frac{\delta^2\Gamma_k[\phi]}{\delta\phi(x)\,\delta\phi(y)}.
\end{equation}


\section{Scale invariance}
\label{sec_scale}

We now discuss how scale invariance emerges in the NPRG formalism. We
first consider a scale-invariant model described by an action
$S_{\rm scal}$.  As we discuss in the following, the existence of such
a model is not necessary for our proof but it is convenient to imagine
that it exists to motivate the form of the expected Ward identity (WI)
of scale invariance.  If such a model existed, this WI would be
obtained by performing the following infinitesimal change of variables
$\varphi(x)\to\varphi(x)+\epsilon(D_\phi+x_\mu\partial_\mu)\varphi(x)$
in the functional integral, with $D_\phi$ the scaling dimension of the
field, usually written in terms of the anomalous dimension $\eta$ as
$D_\phi= (d-2+\eta)/2$. Actually, the analysis of this model and of
its WI faces both UV and infrared (IR) problems. In the IR regime,
non-analyticities are present and, in the UV, it is difficult to
control mathematically the continuum limit: $\Lambda\to\infty$. Let us
first discuss the IR aspect. Since
$\Delta S_k[\phi]$ acts as an IR regulator, the Wilson RG offers a
solution to the IR problem: the regularized model,
Eq. (\ref{eq_part}), is not scale invariant even if the original model
associated with $S_{\rm scal}$ were and, thus, all
$\Gamma_{k>0}^{(n)}(\{p_i\})$ are regular contrarily to
$\Gamma_{k=0}^{(n)}(\{p_i\})$.  The price to pay for regularity is the
breaking of scale invariance that manifests itself through a
modification of the WI [see \cite{Becchi:1996an,Ellwanger:1994iz} for
situations where $R_k$ breaks symmetries].  By enlarging the space of
cutoff functions $R_k$ to arbitrary functions $R_k(x,y)$ that are
neither constrained to satisfy (\ref{cutoff-RG}) nor to be invariant
under rotations and translations [as also done in
\cite{Pawlowski:2005xe}], this modified WI for scale invariance,
obtained from Eqs.~(\ref{eq_part},\ref{eq_legendre}), reads
\begin{align}
  \label{idWardmoddilat-R}
\int_{xy} (D^x+ D^y & +D_R)R_k(x,y)\frac{\delta \Gamma_k}{\delta R_k(x,y)}\nonumber\\
+&
\int_{x} (D^x +D_\phi)\phi(x)\frac{\delta \Gamma_k}{\delta \phi(x)}=0
 \end{align}
 where $D^x=x_\mu\partial_\mu^x$ and $D_R=2d-2D_\phi$ is the scaling
 dimension of $R_k$ which implies that the field renormalization  in 
 Eq.~(\ref{cutoff-RG}) behaves as $Z_k\propto k^{-\eta}$. 
 By constraining $R_k(q)$ to be of  the form
(\ref{cutoff-RG}), Eq.~(\ref{idWardmoddilat-R}) can be conveniently
rewritten  (following~\cite{Ellwanger:1994iz}):
\begin{equation}
\label{idWardmoddilat}
\partial_t \Gamma_k[\phi]=-\int_{x} (D^x +D_\phi)\phi(x)\,\frac{\delta \Gamma_k[\phi]}{\delta \phi(x)}.
\end{equation}
Introducing dimensionless and renormalized quantities (denoted with a tilde) 
\begin{align}
x&=k^{-1} \tilde x\\
\phi(x)&=k^{D_\phi}\tilde \phi(\tilde x)\\ 
\tilde\Gamma_k[\tilde\phi]&=\Gamma_k[\phi]  
\end{align}
Eq.~(\ref{idWardmoddilat}) rewrites:
\begin{equation}
\label{idWardmoddilat2}
\partial_t \tilde\Gamma_k[\tilde\phi]=0.
\end{equation}
Eq.~(\ref{idWardmoddilat2}) means that an hypothetical scale-invariant
action $S_{\rm scal}$ would lead, in its regularized and {\it not}
scale-invariant version $S_{\rm scal}+\Delta S_k$, to a RG flow where
$\tilde\Gamma_k[\tilde \phi]$ would be at a fixed point
$\tilde \Gamma^*[\tilde \phi]$ for all values of $t$:
$\partial_t \tilde \Gamma^*[\tilde \phi]=0$.

The very structure of the Wilson RG (or NPRG) also solves the UV
problem. Actual models have a natural UV cutoff $\Lambda$ (e.g. a
lattice spacing) at which is defined their microscopic action $S$. The
momentum integrals are therefore cut-off at $\Lambda$ and are thus UV finite.
At scales of order $\Lambda$ (i.e., if we
consider correlation functions, in the regime where at least one
external momentum or $k$ or the magnetic field in appropriate units,
is of the order of $\Lambda$), the model is not scale invariant. In
fact scale invariance is an emergent property that appears in the IR
when some parameter (such as the temperature) has been fine-tuned. We call
$S_{\rm crit}$ the corresponding action. In the RG formalism, scale
invariance emerges in the IR regime when integrating the RG flow
starting at $\Gamma_\Lambda[\phi]= S_{\rm crit}[\varphi=\phi] $
because $\tilde\Gamma_k$ gets close to a fixed point for large
negative $t$, that is, $k\ll\Lambda$. As discussed above, the
fixed-point condition coincides with the WI for scale invariance in
presence of a regulator, see
Eqs.~(\ref{idWardmoddilat},\ref{idWardmoddilat2}). As a consequence,
if the RG flow is attracted towards an IR fixed-point, then scale
invariance emerges in the universal, long-distance regime
\footnote{A running
  anomalous dimension can be defined by $\eta_k=-\partial_t \log Z_k$. It is only
  around the fixed point that $\eta_k$ approaches a fixed-point value
  which is nothing else than $\eta$.}. We stress that this discussion
does not rely on the actual existence of a well-defined continuum
limit associated with a scale-invariant action $S_{\rm scal}$ which
is, {\em per se}, an interesting issue, but that we do not need to
address in the present work.

When the microscopic action is slightly different from the
critical one (e.g. choosing a temperature slightly away from the
critical one), the RG trajectory approaches the fixed point and then
stays close to it for a long RG ``time'', before departing.
In this situation, the correlation length $\xi$ is finite but
large and the WI is almost fulfilled for momenta
$\Lambda\gg p\gg\xi^{-1}$. This defines the critical regime of the
theory. The closer the microscopic action is tuned to the critical
one, the larger the correlation length $\xi$, and the better the WI is
fulfilled.

Let us now make two comments. First, when $k\to0$,
$\partial_t \Gamma_k[\phi]$ becomes negligible compared to any
$k$-independent finite scale and $\Gamma_k\to \Gamma$. In this limit,
Eqs.~(\ref{idWardmoddilat},\ref{idWardmoddilat2})  become the usual
WI of scale invariance, as expected. Second, the above analysis shows
that among the continuous infinity of solutions of the fixed point
equation $\partial_t \tilde\Gamma_k[\tilde\phi]=0$, only those that
are regular for all fields are acceptable since they must be the limit
when $k\to 0$ of the smooth evolution of $\tilde\Gamma_k[\tilde\phi]$
from $S_{\rm crit}[\tilde\phi]$.  There is generically only a
discrete, often finite, number of such physical fixed-points.\footnote{Two well-known exceptions to this rule are the line of fixed
points of the $O(2)$ model in $d=2$ \cite{Kosterlitz:1973xp} and the discrete infinity of
(multicritical) $Z_2$-invariant fixed points in $d=2$ \cite{Zamolodchikov:1986db,Morris:1994jc}.}

A characteristic feature of the physical fixed-points is that the
linearized flow around them has a discrete spectrum of eigenvalues
from which some critical exponents can be straightforwardly obtained
\cite{Wilson:1973jj}.  The discrete character of the
eigenperturbations around a fixed point has been studied intensively
by perturbative means.  As for the Wilson RG, it has been studied in
detail in the particular case of the O$(N)$ models in
\cite{Morris:1997xj,Morris:1996nx,Morris:1996xq}.  Although obtained
within the derivative expansion of the exact RG flow (\ref{eqNPRG}),
its discrete character certainly remains true beyond this
approximation. The eigenvalues $\lambda$ are obtained from the flow by
substituting
$\tilde\Gamma_k[\tilde\phi]\to \tilde\Gamma^*[\tilde\phi] +\epsilon\,
\exp(\lambda t)\tilde\gamma[\tilde\phi]$
and retaining only the $O(\epsilon)$ terms. (With our definition of
$t$, a relevant operator has a negative eigenvalue.) This leads to the
eigenvalue problem
\begin{align}
& \lambda\, \tilde\gamma[\tilde\phi]=
 \int_{\tilde{x}} (D^{\tilde{x}} + D_{\phi})\tilde\phi(\tilde{x})\,
\frac{\delta \tilde\gamma}{\delta \tilde\phi(\tilde{x})}\nonumber\\
 &-\frac 1 2\int_{\tilde{x}_i} [(D^{\tilde{x}}+D_R)r(\tilde{x}-\tilde{y})]\,
\tilde{G}_{\tilde{x} \tilde{z}}^*\,\tilde\gamma_{ \tilde{z} \tilde{w}}^{(2)}\,
\tilde{G}_{\tilde{w} \tilde{y}}^*.
\label{flot-perturb}
\end{align}
where $\tilde{x}_i=\{\tilde{x},\tilde{y},\tilde{z},\tilde{w}\}$,
$\tilde{G}^*[\tilde\phi]$ is the dimensionless renormalized propagator
at the fixed point: $\tilde{G}^*=(\tilde\Gamma^{*(2)}+r)^{-1}$ and
$r(\tilde{x})$ is the dimensionless inverse Fourier transform of
$r(q^2/k^2)$ defined in Eq.~(\ref{cutoff-RG}).

We conclude from the above discussion that regularity selects among
all the fixed-point functionals $\tilde\Gamma^*[\tilde\phi]$ those
that are physical, that is, that can be reached by an RG flow from a
physical action $S_{\rm crit}$ and that have a discrete spectrum of
eigenperturbations.


\section{Special conformal transformations}
\label{sec_conformal}

Let us now study conformal invariance by following the same method as
above. As in the case of scale invariance, to motivate the form of the
WI we imagine that a conformally-invariant model, associated with an
action $S_{\rm conf}[\varphi]$ written in terms of a primary field
$\varphi$, exists. We put aside for now the problem of the existence
of this model since, as discussed below, we do not need that it
actually exists for our proof. If such a model existed, the modified
WI would follow from performing the infinitesimal change of variables
$\varphi(x) \to\varphi(x) +\epsilon_\mu (x^2\partial_\mu-2 x_\mu
x_\nu \partial_\nu +2 \alpha x_\mu)\varphi(x)$
in Eq.~(\ref{eq_part}).  By considering general cutoff functions as in
Eq.~(\ref{idWardmoddilat-R}), we find that it reads:
\begin{align}
  \label{ward-conf-modif-R}
  \int_{xy} (K_{\mu}^x-&\, D_R\, x_\mu+ K_{\mu}^y-
                         D_R\, y_\mu)R_k(x,y)\frac{\delta \Gamma_k}{\delta R_k(x,y)}\nonumber\\
  +&
     \int_x (K_{\mu}^x-2 D_\phi x_\mu)\phi(x)\frac{\delta \Gamma_k}{\delta \phi(x)}=0,
\end{align}
with $K_{\mu}^x=x^2\partial_\mu^x - 2 x_\mu x_\nu\partial_\nu^x$. By
specializing $R_k$ to functions of the form Eq.~(\ref{cutoff-RG}) and
requiring again that $Z_k~\propto~k^{-\eta}$,
Eq.~(\ref{ward-conf-modif-R}) can be rewritten
\begin{align}
0=\,\Sigma_k^\mu[\phi]&\equiv\int_{x} (K_{\mu}^x-2 D_\phi x_\mu)\phi(x)\frac{\delta \Gamma_k}{\delta \phi(x)}\nonumber\\
-&
\frac 1 2 \int_{xy}  \partial_t R_k(\vert x-y\vert)\, (x_\mu+y_\mu)\,G_{k,xy}.
\label{ward-conf-modif}
\end{align}
Again, this identity boils down to the usual WI of conformal
invariance in the limit $k\to 0$ where $R_k\to 0$.  

At any fixed point, the scaling dimension of $\Sigma_k^\mu[\phi]$ is fixed by Eq.~(\ref{ward-conf-modif}) to be $-1$. 
We thus define $\tilde\Sigma_k^\mu[\tilde\phi]=k \,\Sigma_k^\mu[\phi]$.
Its flow equation  reads:
\begin{align}
 &\partial_t\tilde\Sigma_k^\mu[\tilde\phi]-\tilde\Sigma_k^\mu[\tilde\phi]=
 \int_{\tilde{x}} (D^{\tilde{x}}+ D_\phi )\tilde\phi(\tilde{x})\,\frac{\delta \tilde\Sigma^\mu_k}{\delta \tilde\phi(\tilde{x})}\nonumber\\
&- \frac 1 2 \int_{\tilde{x}_i} \hspace{-2mm}\left[(D^{\tilde{x}}+D_R)r(\tilde{x}-\tilde{y})\right]
\tilde{G}_{k,\tilde{x} \tilde{z}}\,{\tilde\Sigma^{\mu\;(2)}_{k, \tilde{z} \tilde{w}}}\,\tilde{G}_{k,\tilde{w} \tilde{y}},
\label{flow-sigma}
\end{align}
where $\tilde\Sigma^{\mu\,(2)}_{k}[\tilde\phi]$ is the second functional derivative of $\tilde\Sigma^{\mu}_{k}$.

An important property of $\Sigma_k^\mu$ is that, at the fixed point, it is the integral of
a density with no explicit dependence on $x_\mu$. Indeed, observe that
the left-hand-sides of Eqs.~(\ref{idWardmoddilat-R}) and
(\ref{ward-conf-modif-R}) can be interpreted as the action of the
generators of dilatations $\mathcal D$ and conformal transformations
$\mathcal K_\mu$ on $\Gamma_k$.
\begin{equation}
\begin{split}
\mathcal D \Gamma_k=&\int_{x} (D^x +D_\phi)\phi(x)\frac{\delta \Gamma_k}{\delta \phi(x)}\\
&+\int_{xy} (D^x+ D^y  +D_R)R_k(x,y)\frac{\delta \Gamma_k}{\delta R_k(x,y)}
\end{split}
\end{equation}
\begin{equation}
\begin{split}
\mathcal K_\mu \Gamma_k=&\int_x (K_{\mu}^x-2 D_\phi x_\mu)\phi(x)\frac{\delta \Gamma_k}{\delta \phi(x)}\\
+\int_{xy} (&K_{\mu}^x-\, D_R\, x_\mu+ K_{\mu}^y-
D_R\, y_\mu)R_k(x,y)\frac{\delta \Gamma_k}{\delta R_k(x,y)},
\end{split}
\end{equation}
that is, $\Sigma_\mu=\mathcal K_\mu \Gamma_k$. Similar expressions
can be obtained for the generators of translations $\mathcal P_\mu$
and rotations $\cal J_{\mu\nu}$:
\begin{equation}
\begin{split}
\mathcal P_\mu \Gamma_k=&\int_x\, \partial_\mu\phi(x)\frac{\delta \Gamma_k}{\delta \phi(x)}\\&
+\int_{xy}\, (\partial^x_\mu +\partial^y_\mu)R_k(x,y)\frac{\delta \Gamma_k}{\delta R_k(x,y)}
\end{split}
\end{equation}

\begin{equation}
\begin{split}
\mathcal J_{\mu\nu} \Gamma_k=&\Big[\int_x\, x_\mu \partial_\nu\phi(x)\frac{\delta \Gamma_k}{\delta \phi(x)}\\&
+\int_{xy}\, (x_\mu\partial^x_\nu +y_\mu\partial^y_\nu)R_k(x,y)\frac{\delta \Gamma_k}{\delta R_k(x,y)}\Big]\\
&-[\mu \leftrightarrow\nu]
\end{split}
\end{equation}

It can easily be checked that the generators satisfy the algebra of the conformal group.
In particular, applying $[\mathcal P_\mu,\mathcal K_\nu]=2\delta_{\mu\nu}\mathcal D+2\mathcal J_{\mu\nu}$ to a translation, rotation and
dilatation invariant $\Gamma_k$ yields 
\begin{equation}
  \label{eq_inv_trans_sigma}
\mathcal P_\mu \Sigma_k^\nu=0.  
\end{equation}
Thus, at the fixed point, $\Sigma_k^{\mu }$ is the integral
of a density that does not have an explicit dependence on $x$. This density only depends on 
the field and its derivatives. This proof generalizes trivially
to other scalar models.

\section{A sufficient condition for conformal invariance}
\label{sec_sufficient}

Let us now consider a physical model at criticality. At the scale
$\Lambda$, $\Gamma_{\Lambda}=S=S_{\rm crit}$ and the model is neither
conformally nor scale invariant. However, when $k\ll\Lambda$, the
regularized model gets close to a fixed point and thus
$\tilde G_{k}\simeq \tilde{G}^*$ and
$\tilde\Sigma_{k}^\mu[\tilde\phi]\simeq\tilde\Sigma^{\mu
  *}[\tilde\phi]$.
The key point of our proof is that at the fixed point,
Eq. (\ref{flow-sigma}) is identical to (\ref{flot-perturb}) with
$\tilde{\gamma}[\tilde\phi]\to \tilde\Sigma^{\mu *}[\tilde\phi]$ and
$\lambda= -1$ although these two equations have different physical
meanings.  A {\it sufficient} condition for proving conformal
invariance is therefore to show that there is no integrated vector
eigenperturbation $V_\mu=\int_x \mathcal V_\mu$ of
$\tilde\Gamma^*[\tilde\phi]$ of scaling dimension $D_V=-1$.  If no
such eigenperturbation exists, the conformal WI is satisfied, which
means that the system is conformally invariant at criticality in the long distance regime. Moreover,
the form of the conformal WI (\ref{ward-conf-modif}) fixes the
transformation law fulfilled by $\varphi$, which is the one of a
primary field.


To understand how conformal invariance is related with the scaling
dimension of the vector eigenperturbations, it proves useful to
consider three simple examples. The first one is the Ising model in
$d=4$. The fixed point being gaussian, the eigenvalues are trivially
given by the canonical dimensions. By using the fact that
$\tilde\Sigma^\mu_k$ is $\mathbb{Z}_2-$ and translation-invariant, see Eqs.~(\ref{ward-conf-modif}) and (\ref{eq_inv_trans_sigma}),
we find by inspection that the
vector operator with lowest dimension reads
$\int_x \phi\,\partial_\mu \phi(\partial\phi)^2$. It has therefore
dimension +3.  Note that there exists local vectors operators with
lower scaling dimensions [$\phi\partial _\mu \phi$,
$\phi^3\partial _\mu \phi$ and
$\phi\partial _\mu \phi(\partial_\nu \phi)^2)$]. However these are
total derivatives and are not associated with integrated vector
operators.  In the absence of vector operator of dimension $-1$, we
retrieve the well-known property that this model is conformally
invariant at criticality in the long distance regime \cite{Polchinski:1987dy,Jack:1990eb}.  Using
standard methods, we can compute the corrections to the scaling
dimension of the vector operator
$\int_x \phi\,\partial_\mu \phi(\partial\phi)^2$ in a systematic
expansion in $\epsilon=4-d$. We performed the calculation at one loop
and found that the correction vanishes. The scaling dimension is
therefore $3+\mathcal O(\epsilon^2)$.

This analysis can be extended to the O($N$) models. In $d=4$, there
exists now two independent integrated vector operators,
$\int_x \phi^a(\partial_\mu \phi^a)(\partial_\nu \phi^b)^2$ and
$\int_x \phi^a (\partial_\nu \phi^a)(\partial_\mu \phi^b)(\partial_\nu
\phi^b)$,
with lowest scaling dimension 3.  As in the Ising case, there exist local
operators with lower scaling dimensions, that are however total
derivatives. Our sufficient condition is again fulfilled and we
recover the well-known fact that these models are conformally invariant
in the critical regime for $d=4$.  At one loop the degeneracy of the
scaling dimensions is lifted and we obtain $3+\mathcal O(\epsilon^2)$
and $3-\frac{6\epsilon}{N+8}+\mathcal O(\epsilon^2)$.  

The third example involves a vector field $A_\mu(x)$ and is described
by the (euclidean) action
\begin{equation}
   S=\int_x \left[\frac 1 4\left(\partial_\mu A_\nu -
       \partial_\nu A_\mu \right)^2 + 
     \frac \alpha 2 \left( \partial_\mu A_\mu\right)^2 \right].
     \label{Cardy-model}
\end{equation}
This model is interesting because it is scale invariant but
generically not conformally invariant, except for
$\alpha =\alpha_c=(d-4)/d$ \cite{Riva:2005gd,ElShowk:2011gz}. This
situation can be understood in our context by considering the
contraposition of our sufficient condition, which states that,
assuming that $A_\mu$ is a primary field \cite{ElShowk:2011gz}, a
necessary condition for having scale invariance without conformal
invariance is that there exists an integrated vector operator with
scaling dimension $-1$. It is actually easy to find such an operator
(which is unique, up to a normalization):
$C \int_x A^\mu\left( \partial_\nu A_\nu\right)$. We can understand
the particular case $\alpha=\alpha_c$ by an explicit calculation of
$\Sigma_k^{\mu }[A_\nu]$ from Eq.~(\ref{ward-conf-modif}). This shows
that $C=\alpha d+4-d$ which, as expected, vanishes when
$\alpha=\alpha_c$.

\begin{figure}[t]
  \centering
 \includegraphics[width=.35 \linewidth ]{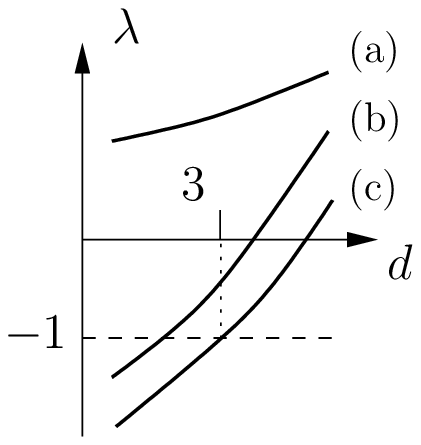}\hspace{1cm}
\includegraphics[width=.35 \linewidth ]{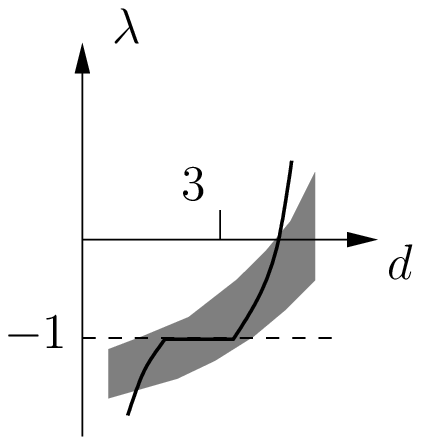}
  \caption{Possible behavior of the lowest eigenvalue $D_V$ associated with a vector perturbation as a function of dimension. Left panel:
  (a) and (b) correspond to typical behavior, (c) to the exceptional case where $D_V=-1$ right in $d=3$. In the three
  cases, conformal invariance holds. Right panel: the shaded area represents a continuum of eigenvalues and the curve
  an eigenvalue $D_V$ having a plateau at $-1$ around $d=3$. In these cases, conformal invariance can be broken.}
  \label{fig_l}
\end{figure}

To conclude this section, we
discuss the plausibility of {\it not} having conformal invariance in $d=3$ for
the O($N$) models at criticality. The only possibility  would be to have a
vector eigenperturbation with eigenvalue $-1$ in this
dimension, see Fig.~\ref{fig_l}.  This would mean that the $d=3$ model
has an integer critical exponent, a property that is highly
improbable. Let us anyhow suppose that one of these eigenvalues
crosses $-1$ right in $d=3$, as in curve (c) of
Fig.~\ref{fig_l}. Then, for any dimension infinitesimally smaller or
larger than three, there would exist no eigenperturbation of dimension
$-1$. The critical system would exhibit conformal symmetry above and
below $d=3$. Since correlation functions of the critical theory are
expected to be continuous functions of $d$, we conclude that, even in
this highly improbable situation, the model would exhibit conformal
invariance at criticality in $d=3$. We are thus led to the more
stringent necessary (but not sufficient) condition: for a critical
model {\it not} to be conformally invariant, there must exist a vector
perturbation of scaling dimension $-1$ in a finite interval of dimensions containing $d=3$.
This could happen either because a discrete eigenvalue is
independent of the dimension in some range of dimension around three
or because there exists a continuum of eigenvalues, see
Fig.~\ref{fig_l}. Such a behavior is, to say the least, not
standard. To our knowledge, this has never been observed in any
interacting model.

The previous arguments are compelling but not mathematically
rigorous. In particular, assuming that the theory can be properly
defined in noninteger dimensions, which is standard but not obvious,
it is hard to control the analytic structure of the critical exponents
in $d$. In the next section, we give a proof in the physically
important case of the Ising model in $d=3$ which does not rely on such
arguments.

\section{Proof in the Ising universality class}
\label{sec_proof}

We concentrate here on the Ising universality class and show that, in
this case, the smallest eigenvalue $D_V$ associated with an integrated
vector perturbation is larger than $-1$ for $d<4$. Using our necessary
condition, this proves that, for the Ising universality class, the
critical regime is conformally invariant.

For simplicity, we consider a lattice version of the
Ginzburg-Landau model whose dynamics is
described by the hamiltonian (or action)
\begin{equation}
  \label{eq:ham_GL}
  S=-J \sum_{\langle ij\rangle}\varphi(i)\varphi(j)+\sum_i U(\varphi(i)).
\end{equation}
where the index $i$ labels the lattice sites, the $\varphi(i)$ take values
in the real domain and $U(\varphi)$ is an even function that diverges
for $|\varphi|\to\infty$. We choose here a hypercubic lattice with
lattice spacing $a$. The original Ising model can be recovered by
considering a potential $U(\varphi)$ strongly peaked around $\varphi=\pm 1$
but the Ginzburg-Landau model which is in the Ising universality class for a
generic potential is more convenient for what follows.

The use of Ginzburg-Landau model
has another advantage. In the case of a quartic potential,
\begin{equation}
 U(\varphi(i))=\frac{r_0}2 \varphi^2(i)+\frac{u_0}{4!} \varphi^4(i)
\end{equation}
and for $d<4$, the model is super-renormalizable.
Its ultraviolet behaviour is therefore controlled by a Gaussian fixed-point. In this case, the existence
of a controlled scaling limit seems to be under control even if, to the best of our knowledge,
there is no mathematical proof of its existence \cite{Rivasseau}. To
compare different models in the Ising universality class we assume below that this
scaling limit does exist in the following precise sense.
Let us consider two local operators $\mathcal{O}_1(i,a)$
and $\mathcal{O}_2(i,a)$.
Let us also introduce (i) a smooth interpolating field $\phi^{\rm interp}(x)$ with $x\in \mathbb{R}^d$
that coincides on the lattice points with $\phi(i)$ and (ii) two interpolating operators ${\cal O}_{1,2}^{\rm interp}(x,a)$
defined by: ${\cal O}_{i}^{\rm interp}={\cal O}_{i}(\phi^{\rm interp})$.
Of course the construction of these interpolating operators is not unique.
We now consider the particular case where ${\cal O}_1$ and ${\cal O}_2$ are such that
${\cal O}_1^{\rm interp}\to {\cal O}_2^{\rm interp}$ when $a\to 0$.
(We notice that this limit is independent of the choice of interpolation used to define $\phi^{\rm  interp}$).
Our assumption, that we call for short ``scaling limit", is that there exists
a multiplicative factor $Z_{\mathcal{O}}(a)$ depending on the lattice spacing such that the correlation
functions of the operators $\mathcal{O}_1(x,a)$ and $Z_{\mathcal{O}}(a)\mathcal{O}_2(x,a)$ are the same for distances much larger than
$a$:
\begin{align}
  \langle \mathcal{O}_1(x,a) &\mathcal{O}_3(y_3) \dots \mathcal{O}_n(y_n)\rangle\nonumber\\
 & \sim 
   Z_{\mathcal{O}}(a) \langle\mathcal{O}_2(x,a)\mathcal{O}_3(y_3) \dots \mathcal{O}_n(y_n)\rangle
\label{continuumlimit}
   \end{align}
where $\{\mathcal{O}_3, \dots, \mathcal{O}_n\}$ are arbitrary local operators and where the equivalence occurs
for $a \ll \min\{|y_3-x|,\dots,|y_n-x|\}$. This hypothesis is a pre-requisite of all Monte Carlo simulations, and
is, of course, satisfied to all orders of perturbation theory in any renormalizable theory. We assume here that
it is also valid nonperturbatively.

Our strategy is to study correlation functions of local vector
operators $\mathcal V_\mu(x)$ and use their critical behavior to find
a bound on the scaling dimension of the {\em integrated} operator
$V_\mu=a^d\sum_i \mathcal V_\mu(i)$. We rightaway mention two
difficulties.

First, there are local vector operators that are total derivatives
and which are therefore not associated with an integrated one. As
discussed before, the vector operator
$\partial_\mu (\varphi^2)$ is such an operator. Note that its scaling
dimension near $d=4$ is lower than that of the vector operators which are not
total derivatives (such as
$\partial_\mu (\varphi)^2(\partial_\nu \varphi)^2$).

Second, operators with the same quantum numbers typically mix together in the
calculation of correlation functions. As a consequence, the critical
behavior of a two-point function of some vector operator is governed
by the lowest scaling dimension of the class of operators with which it mixes.
To be more precise, let us call
$\mathcal V_\mu^{(n)}$ the local vector eigenoperator of scaling
transformations with scaling dimension $D_{\mathcal V}^{(n)}$ (ordered
such that $D_{\mathcal V}^{(0)}\leq D_{\mathcal V}^{(1)}\leq
\cdots$).
The associated two-point correlation function behaves,
in the critical regime, as:
\begin{equation}
  \langle {\mathcal V}_\mu^{(n)}(x){\mathcal V}_\mu^{(n)}(y)\rangle_\co\sim \frac 1{|x-y|^{2D_{\mathcal V}^{(n)}}},
\end{equation}
where the subscript c indicates connected correlation functions,
defined as:
\begin{equation}
\label{eq_co}
  \langle X(x)Y(y)\rangle_\co= \langle X(x)Y(y)\rangle- \langle
  X(x)\rangle\langle Y(y)\rangle
\end{equation}
and where an appropriate normalization has been chosen.  In $d=4$, the
eigenproblem can be solved and the scaling dimensions are the
canonical dimensions. In particular, the eigenoperators with lowest
scaling dimensions are:
\begin{align}
   {\mathcal V}_\mu^{(0,d=4)}&\propto\partial_\mu \phi^2\qquad\qquad\qquad D_{\mathcal V}^{(0,d=4)}=3\\
   {\mathcal V}_\mu^{(1,d=4)}&\propto\partial_\mu \phi^4\qquad\qquad\qquad D_{\mathcal V}^{(1,d=4)}=5\\
   {\mathcal V}_\mu^{(2,d=4)}&\propto\partial_\mu (\partial_\nu \phi)^2\qquad\qquad\, D_{\mathcal V}^{(2,d=4)}=5\\
\label{eq_oper_vec}   {\mathcal V}_\mu^{(3,d=4)}&\propto(\partial_\mu \phi^2)(\partial_\nu \phi)^2\qquad\,\, D_{\mathcal V}^{(3,d=4)}=7\\
&\cdots\nonumber
\end{align}
where $\partial_\mu$ is a lattice discretization of the partial
derivative. (Note that the first three operators are total
derivatives.) In dimension $d<4$, although we do not know the explicit
form of the eigenoperators, we can, in principle, decompose any
vector operator on the basis $\{{\mathcal V}_\mu^{(n)}\}$ and generically, there is a nonvanishing overlap with each of the eigenoperators:
\begin{equation}
   {\mathcal V}_\mu= \sum_n \alpha_n {\mathcal V}_\mu^{(n)}.
\end{equation}
As a consequence, the critical regime of the two-point correlation
function  is dominated by the
smallest critical dimension
\begin{equation}
\label{eq_scalinglower}
  \langle {\mathcal V}_\mu(x) {\mathcal V}_\mu(y) \rangle_\co\sim \frac {\alpha_0^2}{|x-y|^{2D_{\mathcal V}^{(0)}}}.
\end{equation}
We stress that the list of quantum numbers must include those associated with
lattice isometries. For example, we require
scalar (respectively vector) operators defined on the lattice to be even (respectively odd) under parity transformations.

The proof is organized as follows.
Using Lebowitz inequalities \cite{lebowitz74,glimm87},
we derive a lower bound for $D_{\cal V}^{(0)}$
from which follows a lower bound for the scaling dimension of the integrated vector operators.
The proof that the scaling dimension $D_V$ of any integrated vector operator is different
from $-1$ for $d\le4$ is then a direct consequence of this bound.

As a first step, we derive a bound for the correlation function
$\langle \varphi^2(x)\varphi^2(y)\rangle_\co$. We use here the
Lebowitz inequalities \cite{lebowitz74} which state that, considering
two decoupled copies of the ferromagnetic system (that we note
$\varphi$ and $\varphi'$), both described by the action (\ref{eq:ham_GL}),
and considering two sets $A$ and $B$ of lattice points,
\begin{align}
  \label{ineq_leb_def1}
  \begin{split}
 \langle  \prod_{i\in A, j\in B}& [\varphi(i)+\varphi'(i)][\varphi(j)-\varphi'(j)]\rangle\leq\\& 
  \langle \prod_{i\in
  A} [\varphi(i)+\varphi'(i)]\rangle\langle\prod_{j\in B}[\varphi(j)-\varphi'(j)]\rangle   , 
  \end{split}
\\
  \begin{split}
\langle \prod_{i\in A, j\in B}&
  [\varphi(i)+\varphi'(i)][\varphi(j)+\varphi'(j)]\rangle\geq\\& 
 \langle \prod_{i\in
  A} [\varphi(i)+\varphi'(i)]\rangle\langle \prod_{j\in B}[\varphi(j)+\varphi'(j)]\rangle.
\end{split}
\end{align}
In particular, this implies that:
\begin{equation}
  \begin{split}
  \langle(\varphi(x)+\varphi'(x))^2&(\varphi(y)-\varphi'(y))^2\rangle\leq\\
& \langle(\varphi(x)+\varphi'(x))^2\rangle\langle(\varphi(y)-\varphi'(y))^2\rangle.    
  \end{split}
\end{equation}
Expanding the binomials, we readily obtain the following identity:
\begin{equation}
  \langle \varphi^2(x)\varphi^2(y)\rangle_\co\leq 2 G^2(x-y)
\end{equation}
where we have used the fact that the average of an odd number of
fields vanishes for temperatures higher than (or equal to) the
critical temperature. This implies that the connected correlation
function $\langle \varphi^2(x)\varphi^2(y)\rangle_\co$ cannot decrease more
slowly than the square of the propagator at long distances. This
inequality can be generalized to arbitrary even powers of the fields:
\begin{equation}
\label{eq_aleph}
 0\leq \langle \varphi^m(x)\varphi^n(y)\rangle_\co\leq C\, G^2(x-y)
\end{equation}
where $C$ is a positive constant (that depends on  $m$ and $n$) as shown in
the Appendix~\ref{sec_app_bound}.

In the critical regime, scale invariance implies that connected
two-point correlation functions behave as power-laws. In particular:
\begin{equation}
  \label{eq_2ptspower}
  \langle \varphi^m(x)\varphi^n(y)\rangle_\co\sim \frac A{|x-y|^{\aleph_m+\aleph_n}}
\end{equation}
with $A$ a positive constant (see, for example \cite{DiFrancesco:1997nk}). The inequality (\ref{eq_aleph}) implies that
$\aleph_n\geq d-2+\eta$. We can then deduce the asymptotic behavior of
the matrix of second derivatives of this correlation function:
\begin{equation}
  \label{eq_2ptsderpower}
  \begin{split}
    \langle \partial^x_\mu(\varphi^m(x))\partial^y_\nu(\varphi^n(y))\rangle_\co
  \sim & \frac 1{|x-y|^{\aleph_m+\aleph_n+2}}\Big(B \delta_{\mu\nu}\\
&+C\frac{(x-y)_\mu(x-y)_\nu}{(x-y)^2}\Big).
  \end{split}
\end{equation}
where $B$ and $C$ are some constants.

We now consider two local vector operators that are the product of
one power of $\partial_\mu \varphi(x)$ and an odd (finite) number of
fields evaluated at points in a finite neighborhood of $x$:
\begin{align}
  \label{eq_defW1mu}
  &\mathcal W_\mu^{(1)}(x)=\frac12[\partial_\mu\varphi(x)]\sum_{s=\pm 1}\prod_{i=1}^{m-1}\varphi(x+s\, e_i^{(1)}). \\
  \label{eq_defW2mu}
  &\mathcal W_\mu^{(2)}(x)=\frac12[\partial_\mu\varphi(x)]\sum_{s=\pm 1}\prod_{i=1}^{n-1}\varphi(x+s\, e_i^{(2)}). 
\end{align}
where $e_i^{(1)}$ and $e_i^{(2)}$ are some constant lattice vectors\footnote{Since $\Sigma_\mu$, defined in
Eq.~(\ref{ward-conf-modif}), is odd under parity, it is important in what follows to consider only vector operators that are also odd.
This is the reason why the sum over $s$ is necessary in the definitions (\ref{eq_defW1mu}) and (\ref{eq_defW2mu}).}.
The operators $W_\mu^{(1)}(x)$ and $W_\mu^{(2)}(x)$ are, up to a multiplicative constant, other
discretizations of, respectively, the operator $\partial^x_\mu(\varphi^m(x))$ and $\partial^x_\mu(\varphi^n(x))$.
According to the assumption of the existence of the scaling limit, Eq.~(\ref{continuumlimit}), the connected correlation function
$\langle \mathcal W_\mu^{(1)}(x) \mathcal W_\nu^{(2)}(y)\rangle_\co$
has the same asymptotic behavior as in
Eq.~(\ref{eq_2ptsderpower}) up to a multiplicative factor depending on the lattice spacing. Indeed, when $|x-y|$ is much larger than
the lattice spacing $a$, the vectors $e_i^{(1)}$ and $e_i^{(2)}$
can be neglected in (\ref{eq_defW1mu}) and (\ref{eq_defW2mu}) and the local operators $\mathcal W_\mu^{(1)}$ and
$\mathcal W_\mu^{(2)}$ are then proportional to
$\partial_\mu (\varphi^m)$ and $\partial_\nu (\varphi^n)$
respectively as explained before.

Now, any local vector operator ${\mathcal V}_\mu$ on the lattice is a
linear combination of vector operators of the form
(\ref{eq_defW1mu}). For instance, a discretization of the operator
$\partial_\mu(\phi^2)(\partial_\nu\phi)^2$ is given by:
\begin{equation}
\frac {\phi(x)}{16a^3}\left(
    \phi(x+\hat\mu)-\phi(x-\hat\mu)\right)\sum_\nu\left(
    \phi(x+\hat\nu)-\phi(x-\hat\nu)\right)^2
\end{equation}
where the sum runs over all the nearest neighbors of $x$.

Using the triangular inequality, we conclude that:
\begin{align}
  \left|  \langle {\mathcal V}_\mu(x){\mathcal V}_\nu(y)\rangle_\co\right |\leq \frac {Z_{\mu\nu}}{|x-y|^{2(d-1+\eta)}}.
\end{align}
where $Z_{\mu\nu}$ is a positive constant. Using Eq.~(\ref{eq_scalinglower}) this implies that, for all $n$:
\begin{equation}
  D_{\mathcal V}^{(n)}\geq d-1+\eta
\end{equation}
We conclude that the scaling dimension $D_V=D_{\mathcal V}-d$ of any
{\em integrated} vector operator is not smaller than
$-1+\eta$.\footnote{This bound also applies for correlation functions
  of more general operators.} Using the unitarity of
the Minkowskian $\phi^4$ theory, one can prove that $\eta\geq 0$
\cite{ZinnJustin:2002ru} for the Ising universality class. 
Moreover, an interacting massless theory such as the critical Ising model for $d<4$,
has a non-zero $\eta$ \cite{Pohlmeyer:1969av}. As a consequence,
our necessary condition is fulfilled and we
conclude that scale invariance implies conformal invariance for the
Ising universality class for all $d\le4$. 

\section{Conclusions}
\label{sec_conclusion}

Let us now point out some directions of research for the future.  It
is clear that the condition of conformal invariance
(\ref{ward-conf-modif}) can be straightforwardly extended to other
theories (involving scalar, fermionic or vector fields) and it would
be interesting to conclude on the fate of conformal invariance in this
wider class of models. In these systems, it is much more difficult to
find rigorous bounds on correlation functions (that would generalize
the Lebowitz inequalities). It would then prove useful to approach the
problem by computing the scaling behavior of vector operators by
Monte-Carlo simulations.


Another promising line of investigation consists in making use of the
conformal invariance in the Wilson framework to perform actual
calculations of universal quantities. On the one hand, and in the best
case, this would lead to closed and numerically tractable equations
for the critical exponents. On the other hand, the approximation
schemes currently used for solving the Wilson RG flow equation being
incompatible with exact conformal invariance, we can expect that
constraining them to be conformally invariant at the fixed point would improve their
accuracy.

Notice finally that, at first sight, our approach could seem similar
to the one based on the energy-momentum tensor and on the analysis of
the virial current. This is not the case although there is perhaps a
relationship between the two. $\Sigma^\mu_k$ is a functional of $\phi$
and not of $\varphi$; it is built from $\Gamma_k$ and not from $S$;
what matters is that its density vanishes up to a surface term and not
that it is conserved. Moreover, as we already explained, we only deal
with a regularized theory which enables us to consider only the
analytic candidates for $\tilde\Sigma^{\mu *}$ contrary to what should
be done in a nonregularized theory. In any case, a clarification of
the relation between the two approaches would be welcome. In this
respect, our proof of the non-existence of local vector operator of
scaling dimension $d-1$ (conserved or not) might be of interest also
when applied to a hypothetical conserved virial current.


\begin{acknowledgments}
  The authors acknowledge financial support from the ECOS-Sud
  France-Uruguay program U11E01, and from the PEDECIBA. BD and MT also
  thank the Universidad de la Rep\'ublica (Uruguay) for hospitality
  during the completion of this work, and NW the LPTMC for hospitality
  during his sabbatical year 2012-2013. We thank A. Abdesselam, L. Messio, T. Morris, V. Rivasseau
  and V. Rychkov for useful discussions. We thank O. Rosten for pointing out
  the reference \cite{Pohlmeyer:1969av} and V. Rychkov for
  pointing out an error in a previous version of the manuscript.
\end{acknowledgments}

\appendix

\section{Bound for correlation functions $\langle\varphi^m(x)\varphi^n(y)\rangle$}
\label{sec_app_bound}

In this appendix, we derive bounds on the correlation functions
$\langle\varphi^m(x)\varphi^n(y)\rangle_\co$, in the symmetric phase
($T\geq T_c$), with $m$ and $n$ arbitrary integers with the same
parity (otherwise the correlation function vanishes).

We want to show that:
\begin{align}
\label{eq_ineqodd}
  \langle \varphi^a(x)\varphi^b(y)&\rangle_\co\leq C_1 G(x-y)
  \qquad\text{ for odd $a$, $b$ }\\
\label{eq_ineqeven}
  \langle \varphi^a(x)\varphi^b(y)&\rangle_\co\leq C_2 G^2(x-y)
  \hspace{.52cm}\text{ for even $a$, $b$ }
\end{align}
where $C_1$ and $C_2$ are some strictly positive constants and
\begin{equation}
  G(x-y)=\langle\varphi(x)\varphi(y)\rangle.
\end{equation}
Note that for odd $a$ and $b$, the connected and disconnected
correlation functions are equal, see Eq.~\ref{eq_co}.

Property (\ref{eq_ineqodd}) is obvious for $a=b=1$.  The proof of
(\ref{eq_ineqeven}) for $\{a=2,b=2\}$ is presented in the core of the
article. For general $a$ and $b$, the proof is made by
induction. Assuming that the inequalities (\ref{eq_ineqodd}) and
(\ref{eq_ineqeven}) are fulfilled for
$\{a\leq m, b\leq n\}\setminus\{a=m,b=n\}$, we have to prove that
these properties are also valid for $a=m$ and $b=n$.

We first consider the case where $m$ and $n$ are even. Using the
Lebowitz inequality \cite{lebowitz74} [see Eq.~(\ref{ineq_leb_def1})]
\begin{equation}
  \label{eq:Leb1}
  \begin{split}
  \langle[\varphi(x)+\varphi'(x)]^m&[\varphi(y)-\varphi'(y)]^n\rangle\leq\\
& \langle[\varphi(x)+\varphi'(x)]^m\rangle\langle[\varphi(y)-\varphi'(y)]^n\rangle    
  \end{split}
\end{equation}
as well as translation invariance and the binomial expansion, we obtain:
\begin{equation}
  \label{eq:Leb2}
  \begin{split}
  \sum_{a=0}^m\sum_{b=0}^n&(-1)^b{m\choose a}{n\choose b}\Big[
\langle\varphi^a(x)\varphi^b(y)\rangle_\co\langle\varphi^{m-a}(x)\varphi^{n-b}(y)\rangle_\co
\\&+\langle\varphi^a(0)\rangle\langle\varphi^b(0)\rangle\langle\varphi^{m-a}(x)\varphi^{n-b}(y)\rangle_\co
\\&+\langle\varphi^a(x)\varphi^b(y)\rangle_\co\langle\varphi^{m-a}(0)\rangle\langle\varphi^{n-b}(0)\rangle\Big]
\leq  0    
  \end{split}
\end{equation}
Writing explicitly the terms with $a\in\{0,m\}$ and $b\in\{0,n\}$, we
get the following bound:
\begin{equation}
  \label{eq:Leb3}
  \begin{split}
2    \langle
\varphi^m(x)&\varphi^n(y)\rangle_\co\leq\sum_{a\in\{1,3,\cdots,m-1\}}\sum_{b\in\{1,3,
  \cdots,n-1\}}{m\choose a}{n\choose b}\times\\
&\langle\varphi^a(x)\varphi^b(y)\rangle_\co\langle\varphi^{m-a}(x)\varphi^{n-b}(y)\rangle_\co
  \end{split}
\end{equation}
where we have used the fact that connected correlation functions, as
well as $\langle\varphi^m(0)\rangle$ are non-negative
\cite{Griffiths67,Kelly68} to eliminate the terms with even $b$.  We
observe that there appears in the right-hand side only connected
correlation functions with at most $m-1$ powers of $\varphi(x)$ and at
most $n-1$ powers of $\varphi(y)$. By hypothesis, properties
(\ref{eq_ineqodd}) and (\ref{eq_ineqeven}) are thus valid for these
correlation functions. Furthermore, we observe that the right-hand
side involves a sum of terms that are
a product of two correlation functions with odd powers of the
fields. In both cases, by using properties (\ref{eq_ineqodd}) and
(\ref{eq_ineqeven}), this quantity is bounded by some positive
constant times $G^2(x-y)$.\footnote{We use here the fact that, on the
  lattice, $G$ is bounded by some positive constant
  ($G(r)<D$). Consequently, we can further use the bound
  $G^{2n}(r)\leq C G^{2}(r)$ where $n$ is a positive integer and $C$ a
  positive constant.} This concludes the proof of the induction hypothesis in the case of even $m$ and $n$.

We now turn to the case where $m$ and $n$ are odd. We now make use of
the Lebowitz inequality
\begin{equation}
  \label{eq:Leb4}
  \begin{split}
    \langle[&\varphi(x)+\varphi'(x)]^{m-1}[\varphi(x)-\varphi'(x)][\varphi(y)-\varphi'(y)]^n\rangle\leq\\
    &
    \langle[\varphi(x)+\varphi'(x)]^{m-1}\rangle\langle[\varphi(x)-\varphi'(x)][\varphi(y)-\varphi'(y)]^n\rangle
  \end{split}
\end{equation}
which is of interest if $m>1$ (we can obviously derive a similar
inequality with $\{m,x\}\leftrightarrow\{n,y\}$ which can be applied
in the case $m=1$). We again use the binomial expansion and the
positivity of (connected and disconnected) correlation functions to
obtain the following inequality: 
\begin{equation}
  \label{eq:Leb5}
  \begin{split}
\langle
\varphi^m&(x)\varphi^n(y)\rangle_\co\leq\\&\sum_{a\in\{0,2,\cdots,m-1\}}\sum_{b\in\{1,3,
  \cdots,n-1\}}{m-1\choose a}{n\choose b}\times\\
&\langle\varphi^a(0)\rangle\langle\varphi^{m-a-1}(0)\rangle\langle\varphi(x)\varphi^{b}(y)\rangle_\co\langle\varphi^{n-b}(0)\rangle\\
&+\sum_{a\in\{1,3,\cdots,m-2\}}\sum_{b\in\{0,2,
  \cdots,n\}}{m-1 \choose a}{n\choose b}\times\\
&\langle\varphi^{a+1}(x)\varphi^b(y)\rangle\langle\varphi^{m-1-a}(x)\varphi^{n-b}(y)\rangle
  \end{split}
\end{equation}
The first term involves the product of a correlation function with odd
powers of the fields and a positive constant. The second sum involves
either a product of two correlation functions, one with even and one
with odd powers of the fields, or the product of a correlation
function with odd powers of the fields and a positive constant. In all
cases, the correlation functions that appear in the right-hand side
fulfill the conditions of our hypothesis. We therefore conclude that
$\langle \varphi^m(x)\varphi^n(y)\rangle_\co$ satisfies property
(\ref{eq_ineqodd}) for $m$ and $n$ odd (see Footnote 6). This concludes the proof of the induction hypothesis.

Using the fact that the property (\ref{eq_ineqodd}) is valid for
$a=b=1$, it is easy to check, by applying several times the induction
property, that (\ref{eq_ineqodd}) and (\ref{eq_ineqeven}) are valid
for any $a$ and $b$.


\begin{thebibliography}{0}
\expandafter\ifx\csname natexlab\endcsname\relax\def\natexlab#1{#1}\fi
\expandafter\ifx\csname bibnamefont\endcsname\relax
  \def\bibnamefont#1{#1}\fi
\expandafter\ifx\csname bibfnamefont\endcsname\relax
  \def\bibfnamefont#1{#1}\fi
\expandafter\ifx\csname citenamefont\endcsname\relax
  \def\citenamefont#1{#1}\fi
\expandafter\ifx\csname url\endcsname\relax
  \def\url#1{\texttt{#1}}\fi
\expandafter\ifx\csname urlprefix\endcsname\relax\def\urlprefix{URL }\fi
\providecommand{\bibinfo}[2]{#2}
\providecommand{\eprint}[2][]{\url{#2}}

\end{thebibliography}


\vspace{1.5cm}

\begin{thebibliography}{10}

\bibitem{Maldacena:1997re}
  J.~M.~Maldacena,
  Int.\ J.\ Theor.\ Phys.\  {\bf 38}, 1113 (1999)
  [Adv.\ Theor.\ Math.\ Phys.\  {\bf 2}, 231 (1998)].

\bibitem{ElShowk:2012ht}
  S.~El-Showk, M.~F.~Paulos, D.~Poland, S.~Rychkov, D.~Simmons-Duffin and A.~Vichi,
  Phys.\ Rev.\ D {\bf 86}, 025022 (2012).

\bibitem{Gliozzi:2014jsa}
  F.~Gliozzi and A.~Rago,
  J. High Energy Phys.  { 10}  (2014) 042.

 
\bibitem{El-Showk:2014dwa}
  S.~El-Showk, M.~F.~Paulos, D.~Poland, S.~Rychkov, D.~Simmons-Duffin and A.~Vichi,
  J.\ Stat.\ Phys. {\bf 157}, 869 (2014).
  
\bibitem{Kos:2013tga}
  F.~Kos, D.~Poland and D.~Simmons-Duffin,
  J. High Energy Phys. {06} (2014) 091.
 


\bibitem{El-Showk:2013nia}
  S.~El-Showk, M.~Paulos, D.~Poland, S.~Rychkov, D.~Simmons-Duffin and A.~Vichi,
  Phys.\ Rev.\ Lett.\  {\bf 112}, 141601 (2014).
  
  
\bibitem{Nakayama:2014sba} 
  Y.~Nakayama and T.~Ohtsuki,
  Phys. Rev. D {\bf 91}, 021901(R) (2015).
  
\bibitem{Virasoro:1969zu}
  M.~S.~Virasoro,
  Phys.\ Rev.\ D {\bf 1}, 2933 (1970).

\bibitem{Polyakov:1970xd}
  A.~M.~Polyakov,
  JETP Lett.\  {\bf 12}, 381 (1970)
  [Pisma Zh.\ Eksp.\ Teor.\ Fiz.\  {\bf 12}, 538 (1970)].



\bibitem{Migdal:1972tk}
  A.~A.~Migdal,
  Phys.\ Lett.\ B {\bf 37}, 386 (1971).

\bibitem{Belavin:1984vu}
  A.~A.~Belavin, A.~M.~Polyakov and A.~B.~Zamolodchikov,
  Nucl.\ Phys.\ B {\bf 241}, 333 (1984).

\bibitem{Zamolodchikov:1986gt}
  A.~B.~Zamolodchikov,
  JETP Lett.\  {\bf 43}, 730 (1986)
  [Pisma Zh.\ Eksp.\ Teor.\ Fiz.\  {\bf 43}, 565 (1986)].

\bibitem{Polchinski:1987dy}
  J.~Polchinski,
  Nucl.\ Phys.\ B {\bf 303}, 226 (1988).

\bibitem{DiFrancesco:1997nk}
  P.~Di Francesco, P.~Mathieu and D.~S\'en\'echal,
 {\it Conformal field theory}
  (Springer, New York, 1997), p. 890.

\bibitem{Wilson:1973jj}
  K.~G.~Wilson and J.~B.~Kogut,
  Phys.\ Rept.\  {\bf 12}, 75 (1974).

  
\bibitem{Jack:1990eb}
  I.~Jack and H.~Osborn,
  Nucl.\ Phys.\ B {\bf 343}, 647 (1990).

\bibitem{Luty:2012ww}
  M.~A.~Luty, J.~Polchinski and R.~Rattazzi,
  J. High Energy Phys. {01} (2013) 152.

\bibitem{Dymarsky:2013pqa}
  A.~Dymarsky, Z.~Komargodski, A.~Schwimmer and S.~Theisen,
  arXiv:1309.2921 [hep-th].

\bibitem{Dymarsky:2014zja}
  A.~Dymarsky, K.~Farnsworth, Z.~Komargodski, M.~A.~Luty and V.~Prilepina,
  arXiv:1402.6322 [hep-th].




\bibitem{Schafer:1976ss}
  L.~Schafer,
  J.\ Phys.\ A {\bf 9}, 377 (1976).

\bibitem{Rosten:2014oja}
  O.~J.~Rosten,
  arXiv:1411.2603 [hep-th].
  
\bibitem{Polchinski:1983gv}
  J.~Polchinski,
  Nucl.\ Phys.\ B {\bf 231}, 269 (1984).

\bibitem{Wetterich:1992yh}
  C.~Wetterich,
  Phys.\ Lett.\ B {\bf 301}, 90 (1993).
  
\bibitem{Ellwanger:1993kk} 
  U.~Ellwanger,
  Z.\ Phys.\ C {\bf 58}, 619 (1993).

\bibitem{Morris:1993qb} 
  T.~R.~Morris,
  Int.\ J.\ Mod.\ Phys.\ A {\bf 9}, 2411 (1994).
  
\bibitem{Berges}
J. Berges, N. Tetradis and  C. Wetterich, Phys.Rept. {\bf 363}, 223 (2002).

\bibitem{Machado:2010wi}
  T.~Machado and N.~Dupuis,
  Phys.\ Rev.\ E {\bf 82},  041128 (2010).
  
\bibitem{Becchi:1996an}
  C.~Becchi,
  hep-th/9607188.
  
  


\bibitem{Ellwanger:1994iz}
  U.~Ellwanger,
  Phys.\ Lett.\ B {\bf 335}, 364 (1994).

\bibitem{Pawlowski:2005xe} 
  J.~M.~Pawlowski,
  Annals Phys. (N.Y.) \  {\bf 322}, 2831 (2007).
 

\bibitem{Kosterlitz:1973xp}
  J.~M.~Kosterlitz and D.~J.~Thouless,
  J.\ Phys.\ C {\bf 6}, 1181 (1973).
  

\bibitem{Zamolodchikov:1986db}
  A.~B.~Zamolodchikov,
  Sov.\ J.\ Nucl.\ Phys.\  {\bf 44}, 529 (1986)
  [Yad.\ Fiz.\  {\bf 44}, 821 (1986)].

\bibitem{Morris:1994jc}
  T.~R.~Morris,
  Phys.\ Lett.\ B {\bf 345}, 139 (1995)


\bibitem{Morris:1997xj}
  T.~R.~Morris and M.~D.~Turner,
  Nucl.\ Phys.\ B {\bf 509}, 637 (1998).

\bibitem{Morris:1996nx}
  T.~R.~Morris,
  Phys.\ Rev.\ Lett.\  {\bf 77}, 1658 (1996).

\bibitem{Morris:1996xq} 
  T.~R.~Morris,
  Nucl.\ Phys.\ B {\bf 495}, 477 (1997).
  
\bibitem{Riva:2005gd}
  V.~Riva and J.~L.~Cardy,
  Phys.\ Lett.\ B {\bf 622}, 339 (2005).
  
\bibitem{ElShowk:2011gz} 
  S.~El-Showk, Y.~Nakayama and S.~Rychkov,
  Nucl.\ Phys.\ B {\bf 848}, 578 (2011)


\bibitem{Rivasseau}
A. Abdesselam and V. Rivasseau, private communication.

\bibitem{lebowitz74}
J.L. Lebowitz, Commun. Math. Phys. {\bf 35}, 87 (1974).


\bibitem{glimm87}
J. Glimm and A. Jaffe, {\it Quantum Physics: A Functional  Integral Point of View}, 
2nd ed. (Springer-Verlag, New York, 1987).


\bibitem{ZinnJustin:2002ru} 
  J.~Zinn-Justin,
  Int.\ Ser.\ Monogr.\ Phys.\  {\bf 113}, 1 (2002).
  

\bibitem{Pohlmeyer:1969av} 
  K.~Pohlmeyer,
  Commun.\ Math.\ Phys.\  {\bf 12}, 204 (1969).
  
  
  
\bibitem{Griffiths67}
R. B. Griffiths, J. Math. Phys. {\bf 8}, 478 (1967); {\bf 8}, 484 (1967).

\bibitem{Kelly68}
 D. G. Kelly and S. Sherman, J. Math. Phys. {\bf 9}, 466 (1968).








  
\end{thebibliography}
\end{document}